\documentclass[twocolumn,
prd,showpacs,floats,amsmath,amssymb]{revtex4}
\usepackage{epsfig}
\usepackage{amsmath}
\renewcommand{\(}{\left(}
\renewcommand{\)}{\right )}

\def\fslash#1{#1 \!\!\! \slash}
\def\beq{\begin{equation}}
\def\eeq{\end{equation}}
\def\pa{\partial}

\def\varp{\varepsilon}
\def\bea{\arraycolsep .1em \begin{eqnarray}}
\def\eea{\end{eqnarray}}

\def\vp{{\bf p}}

\def\vk{{\bf k}}

\def\Tr{{\rm Tr}}

\let\be=\beta
\let\no=\nonumber

\def\eq#1{Eq.(\ref{#1})}
\def\refr#1{\cite{#1}}

\def\s0#1#2{\mbox{\small{$ \frac{#1}{#2} $}}}
\def\0#1#2{\frac{#1}{#2}}

\def\plb#1#2#3{#2 Phys. Lett. {\bf B #1}, #3 (#2)}
\def\ctp#1#2#3{#2 Commun.\ Theor.\ Phys. \ {\bf #1}, #3 (#2)}

\def\pra#1#2#3{#2 Phys. Rev.  {\bf A #1}, #3 (#2)}
\def\prb#1#2#3{#2 Phys. Rev.  {\bf B #1}, #3 (#2)}
\def\prc#1#2#3{#2 Phys. Rev.  {\bf C #1}, #3 (#2)}

\def\prl#1#2#3{#2 Phys. Rev. Lett. {\bf #1}, #3 (#2)}

\def\ann#1#2#3{#2 Ann. Phys. (N.Y.) {\bf #1}, #3 (#2)}
\def\anp#1#2#3{#2 Adv. Nucl. Phys. {\bf #1}, #3 (#2)}

\def\ijmpb#1#2#3{#2 Int.\ J.\ Mod.\ Phys.\ {\bf B #1}, #3 (#2)}

\def\ibid#1#2#3{#2 {\it ibid.}, {\bf #1}, #3 (#2)}

\def\epl#1#2#3{#2 Europhys.\ Lett.{\bf #1}, #3 (#2)}
\def\sci#1#2#3{#2 Science {\bf #1}, #3 (#2)}
\begin{document}
\title{
Ground state energy of unitary fermion gas with the Thomson
Problem approach}
\author{Ji-sheng Chen \footnote{chenjs@iopp.ccnu.edu.cn}}
\affiliation{
Institute of Particle Physics $\&$ Physics Department, Central
China Normal University, Wuhan 430079, People's Republic of China}

\begin{abstract}
The dimensionless universal coefficient $\xi$ defines the ratio of
the unitary fermions energy density to that for the ideal
non-interacting ones in the non-relativistic limit with $T=0$. The
classical Thomson Problem is
    taken as a nonperturbative quantum many-body arm to address the
    ground state energy including the low energy nonlinear quantum fluctuation/correlation
    effects.
With the relativistic Dirac continuum field theory formalism, the
concise expression for the energy density
    functional of the strongly interacting limit fermions at both finite temperature
    and density is obtained.
Analytically, the universal factor is calculated to be $\xi
=\frac{4}{9}$. The energy gap is $\Delta
=\frac{5}{18}{k_f^2}/(2m)$.\end{abstract}
\maketitle
With the further developments of the Bardeen-Cooper-Schrieffer
theory, the possibility about the existence of the fermions
superfluidity in the dilute gas system motivates widely
theoretical studies and experimental efforts. Since DeMarco and
Jin achieved the Fermi degeneracy\refr{DeMarco}, the ultra-cold
fermion atoms gas has stirred intense interest about the
fundamental Fermi-Dirac statistical physics in the strongly
interacting limit.

Across the Feshbach resonance regime, the interaction changes from
weakly to strongly attractive according to the magnitude of the
magnetic field. At the midpoint of this crossover unitary regime
from Bardeen-Cooper-Schrieffer(BCS) to Bose-Einstein
    condensation(BEC), the
scattering length will diverge due to the existence of a
zero-energy bound state for the two-body system. In this limit,
the only dimensionful parameter is the Fermi momentum $k_f$ at
$T=0$. The corresponding energy scale is the Fermi kinetic energy
    $\varepsilon_f={k_f^2}/{(2m)}$,
    while $m$ is the fermion mass.
According to dimensional analysis, the system details do not
contribute to the
    thermodynamics properties,
    i.e., the thermodynamics properties are universal
    \refr{Bertsch1999,Baker2001,Heiselberg2000,physics/0303094,Ho2004,Astrakharchik2004,Astrakharchik2005,Schwenk2005,Carlson2005,Cohen2005,Dean2006,Chevy2006,Hu2006,Nishida2006,Bhattacharyya2006,Rupak2006,cond-mat/0404687,Bulgac2005}.
The energy density should be proportional to that of a free Fermi
    gas
$\label{universal}
   E/V = \xi\(E/V\)_{free} =\xi \035  n
    \epsilon_f$.

This fundamental dimensionless universal constant $\xi$ has
attracted much attention
    theoretically/experimentally in recent years. Various
    theoretical approaches have been tried and the results
    differ from each other remarkably, for example, see Ref.\refr{Bertsch1999,Baker2001,Heiselberg2000,physics/0303094,Ho2004,Astrakharchik2004,Astrakharchik2005,Schwenk2005,Carlson2005,Cohen2005,Dean2006,Chevy2006,Hu2006,Nishida2006,Bhattacharyya2006,Rupak2006,cond-mat/0404687,Bulgac2005}
    and references therein.
    The theoretical calculations are about $\xi \sim 0.3-0.6$.
Intriguingly, the experimental results are also quite
    different from each other, for example, $\xi \approx 0.74 \pm 0.07$\refr{Gehm2003},
    $\xi \approx 0.7$\refr{Bourdel2003},
    $\xi=0.32^{+0.13}_{-0.10}$\refr{Bartenstein2004},
    $\xi=0.51\pm 0.04$\refr{Kinast2005},
    $\xi=0.46\pm 0.05$\refr{Partridge2006}, $\xi=0.46^{+
    0.05}_{-0.12}$\refr{Stewart2006}.
The recent lattice result is
    $\xi=0.25\pm 0.03$\refr{Lee2006}.

How to approach the exact value of $\xi$ analytically is a
    bewitching topic in the Fermi-Dirac statistical physics.
To attack this intriguing topic is a seriously
    difficult problem in many-body theory. The essential task
is how to incorporate the
    nonlinear quantum fluctuation/correlation effects into the
    thermodynamics by going beyond any naive loop diagram expansions or the lowest
    order mean field theory.
To our knowledge, the hitherto considerations looking for $\xi$
have been solely limited in the non-relativistic frameworks and
with quite different results. How about a relativistic Dirac
phenomenology attempt?

{\bf\underline{Motivation}}: Essentially, the
    unitary physics with infinite scattering lengths is quite
    similar to the \textit{universal} strongly instantaneous Coulomb correlation thermodynamics in a compact nuclear confinement
 environment resulting from the competition of long and short range
    forces\refr{chen2005}. At the
crossover point, the
    cross-section between the two-body particle is limited by $\sigma \sim
    {4\pi}/{k^2}$ ($k$ is the relative wavevector of the colliding particles),
    while the gauge vector boson propagator is $\Delta _{\mu\nu}\sim {g_{\mu\nu}}/{k^2}$ in the gauge
    field theory.
The former is short range but exact long range infrared
    correlation while the latter is long range one. The analogism motivates us to use
    the latter to model the former.

In this context, it is also instructive to recall the intermediate
vector boson (IVB) hypothesis in weak interaction theory. The
local intermediate vector boson theory is related with the
current-current(CC) contact interaction version through the
corresponding connection: ${g^2_W}/{m_W^2}\equiv \0G{\sqrt{2}}$.
In the low energy limit $k\rightarrow 0$, the two IVB and CC
theories are identical. To model the unitary limit, we ``let"
${g^2_W}/{m_W^2}\rightarrow \infty$ with an arbitrary large charge
$g_W$ or with an arbitrary small mass gap $m_W$.

It is usually assumed that there exists a uniform opposite charged
background to ensure the stability in discussing the
    electrons system thermodynamics. This is the well-known classical Thomson
    Problem\refr{Thomson}. Of course, there is not a finished standard answer for itself/generalized version for over a
century\refr{Bowick,Luca2005}. Furthermore, a consummate method to
gauge the infrared
    singularity in the dense and hot gauge theory is still to be
    looked for.
However, we will turn the unitary limit topic into the same
    involved infrared problem.
Then, the Thomson Problem is used as a potential quantum
    many-body nonperturbative arm to attack the infrared one.

To achieve the intriguing physics with the unusual in-medium
relativistic Lorentz invariance breaking at unitarity, an infrared
correlating
    ``QED" Lagrangian is proposed in this {\it Letter}.
Let the fermion have an ``electric" charge $g$ in addition to
other internal global $U(1)$ symmetry quantum
    numbers.
According to the general stability principle, the system should be
stabilized by a fictive uniform opposite charged
    background in the meantime. This particular assumption makes it possible for us to deal with the challenging infinity.

The natural units $c=\hbar =k_B=1$ are used.

To perform the path integral presentation as a nonperturbative
starting point, the considered effective actions involve the
interaction of
    Dirac fermions with an auxiliary Proca-like Lorentz vector boson field\refr{Proca,Jackson}
\bea
    {\cal L}_{matter}=&&{\bar \psi} (i \gamma _\mu\pa ^\mu -m) \psi
    ;\no\\
    {\cal L}_{A, free}=&& -\014 F_{\mu\nu} F^{\mu\nu};\no\\
   \label{electron-current}
    {\cal L}_{I, A-\psi}=&&\012 m_{\mbox{background}}^2 A_\mu A^\mu +A_\mu J^\mu,
\eea
    where $m$ is the bare fermion mass.
The $A_\mu$ is the vector
    field with the field stress
\bea
    F_{\mu\nu}=\pa_\mu A_\nu-\pa_\nu A_\mu.
\eea

In the action ${\cal L}_{I, A-\psi}$, the electric vector current
    contributed by the fermions is
\bea
    J^\mu = g {\bar \psi} \gamma ^\mu \psi.
\eea Based on the local gauge invariant free Lagrangian,
    the many-body interactions can be introduced with a hidden local
    symmetry (HLS) manner\refr{peskin1995,Dvali2005}
\bea
    {\cal L}_I=-\014 F_{\mu\nu}F^{\mu\nu}+|D_\mu H|^2+V(H)+A_\mu J^\mu.
\eea

The many-body Lorentz violation environment modulates the bare
    two-body interaction.
The vector boson mode mass gap $m_{\mbox{background}}$ is not an
    internal degree of freedom.
It appears as a free parameter and
    controls the interaction strength between the particles.
It indicates the interaction of the opposite electric charged
    background or the stochastic  many-body potential caused
    by the strongly fluctuating/correlating effects.
Below, the suffix ``background" will be emitted for brevity just
    with a tilde symbol $m_{\tilde{B}}$.
The un-physical coupling constant $g$ also represents the unitary
    limit/infrared characteristic ``two-body" bare potential.

The electric current conservation/gauge invariance is guaranteed
by the Lorentz transversality
     condition with HLS formalism
\bea
    \pa _\mu A^\mu=0,
\eea
 which can be naturally realized by taking the relativistic Hartree instantaneous approximation (RHA).

In terms of the functional path
    integral\refr{walecka1974,serot1986,chen2005,kapusta1989},
    the auxiliary effective potential reads
\bea
   \label{potential}\Omega/V
       &&=-\012 m_{\tilde{B}}^2 A_0 ^2 -2 T  \int _k
        \left [\ln (1+e^{-\beta (E_k-\mu^*)})\right.\no\\
       &&
       \left.~~~~~~~~ +\ln (1+e^{-\beta (E_k+\mu^*)})\right ],
\eea where $``2"$ represents the (hyperfine-)spin projection of
the fermions with $\int _k=\int {d^3{\bf k}}/{(2\pi )^3}$ and $\be
=1/T$ being the inverse temperature. The tadpole diagram with the
boson self-energy for the full
    fermion propagator leads to
\bea\label{electron-field}
        A_0=-\0{g}{m_{\tilde{B}}^2}n,
\eea
    with the fermions (electric charge number) density defined by
    the thermodynamics relation
    ${\pa \Omega}/{\pa \mu}|_{A_0}\equiv-n$
\bea
    n=2\int _k \left [f(\mu ^*, T )-\overline{f}(\mu ^*, T )\right ].
\eea In the above expressions, \bea
     f(\mu ^*, T )=\01{e^{\be (E_k-\mu ^*)}+1},~~~~
    \overline{f}(\mu ^*, T )=\01{e^{\be (E_k+\mu ^*)}+1},\no
\eea
    are the distribution functions for (anti-)particles with $E_k =\sqrt{{\bf k}^2+m^2}$.
From \eq{electron-field}, the effective chemical potential
    $\mu ^*$ is defined with a gauge invariant manner
\bea\label{chemical}
    \mu^*\equiv\mu+\mu _{I}
    =\mu -\0{g^2}{m_{\tilde{B}}^2 } n,
\eea
    where $\mu$ is the global chemical potential.
The spirit is quite similar to  the Kohn-Sham density functional
    theory\refr{Kohn1965}.

Using \eq{potential} and with the
    thermodynamics relation \bea
    \epsilon =\01V \0{\pa (\beta \Omega )}{\pa \beta }+\mu  n,\eea
    one obtains the energy density functional
\bea\label{energy}
    \epsilon =\012\0{g^2}{m_{\tilde{B}}^2} n^2 +2\int_k E_k \left [f (\mu ^*,T)+{\bar f}(\mu ^*,T)\right
    ].
\eea
    The second term in \eq{energy} appears as very much the analytical formalisms for the free Fermi-Dirac gas.
However, the correlating effects are implicitly included through
    the effective chemical potential esp. for $T\neq 0$.


The mass gap parameter $m_{\tilde{B}}^2$ is a Lagrange
\textit{multiplier} that enforces relevant constraints and
reflects quantum fluctuating effects consequently. The remaining
central task is how to determine the unknown many-body
    stochastic potential characterized by the coupling constants
    $m_{\tilde{B}}^2$ and $g^2$. The answer can be found from the auxiliary effective
    potential, i.e., the composite system should be ``charge neutralized" through the
fictive background with the artificial conditional extremum
\refr{gulminelli2003}\bea\label{stability}
     \0{\pa\Omega }{\pa \mu}|_{m_{\tilde{B}}^2,T}=0.
\eea With \eq{stability}, one can
have\bea\label{debye-mass-relation}
    m_{\tilde{B}}^2&&=-\0{g^2}{\pi^2} \int _0^\infty d|\vk|
    \0{(2\vk ^2+m^2)}{E_k}\left [f (\mu ^*,T)+{\bar f}(\mu ^*,T)\right ]\no\\
    &&\equiv-m_D^2,
\eea which is the \textit{negative} of the gauge invariant
    Debye (Thomas-Fermi)
    screening mass squared $m_D^2$, i.e., there is a
    minus sign between $m_{\tilde{B}}^2$ and $m_D^2$.
The Debye mass parameter can be also directly calculated with the
    vector boson polarization through the Dyson-Schwinger
    equation (relativistic random phase
    approximation-RPA)\refr{chen2002}
\bea\label{self-energy}
    \Pi ^{\mu\nu}_{A}(p_0,\vp)=g^2T\sum _{k_0}\int_k \Tr
        \left [\gamma ^\mu \0{1}{\fslash{k}-m}\gamma
        ^\nu\0{1}{(\fslash{k}-\fslash{p})-m}\right ],\no\\
\eea
   with the full fermion propagator \eq{chemical} by noting
\bea
   m_D^2=-\Pi^{00}_A(0,|\vp|\to 0).
\eea
     In \eq{self-energy}, the 0-component of the four-momentum
    $k=(k_0,\vk )$ in the fermion loop is related to temperature T and
    effective chemical potential
    $\mu ^*$ via $k_0=(2 n+1) \pi T i +\mu ^*$. It is very
    interesting that the Thomson stability condition can give the
    gauge invariant Debye mass, which is \textit{exactly} consistent with
    the Dyson-Schwinger equation.

At $T=0$, the parameter $m_{\tilde{B}}^2$ is \bea\label{dos}
    m_{\tilde{B}}^2
    =-\0{g^2}{\pi ^2}k_fE_{f},
\eea
    with $k_f$ being the Fermi momentum and
    $E_f=\sqrt{k_f^2+m^2}$ the relativistic Fermi kinetic energy.

The \eq{energy} with \eq{debye-mass-relation}
    ($m^2_{{\tilde B}}=-m_D^2$) are our main results, from which one
        can further study the strongly correlating
        fermions thermodynamics in the unitary limit.
The collective interaction contribution is \textit{negative} for
the
    physical energy density functional.
Especially, there is not any remained parameter because the
    un-physical coupling constant $g$
    appears simultaneously in the denominator and numerator within the
    relevant analytical expressions through $m_{\tilde{B}}^2$.
From \eq{dos}, one can see that the magnitude of this fraction
ratio $m_{\tilde{B}}^2/g^2$ characterize the density of states and
consequently the fluctuating contributions.


With the mathematically well defined energy density functional
\eq{energy} and with \eq{debye-mass-relation}, we now return to
the final discussions. At $T\to 0$, the
    energy density is
\bea \epsilon =-\0{k_f^5}{18\pi^2 E_f }+
\0{(2k_f^2+m^2)k_fE_f-m^4\ln\0{k_f+E_f}{m}}{8\pi^2}. \eea In the
non-relativistic limit $k_f/m\ll 1$, one can expand
    $\epsilon $ according to the Taylor series of $k_f/m$\bea \epsilon =m n
    +\04{15}n\varepsilon_f+\05{42}n \0{\varepsilon _f^2}{m}-\013n
    \0{\varepsilon _f^3}{m^2}+\cdots,
\eea
    with the Fermi kinetic energy $\varp_f ={k_f^2}/{(2m)}$ and
    particle number density
    $n={k_f^3}/{(3\pi^2)}$.
Therefore, one can obtain the ground state binding energy (energy
    per particle)
\bea
    e_b=\0{\epsilon}{n}-m=\(\04{15}+\05{42}\0{\varepsilon_f}{m}-\013\(\0{\varepsilon_f}{m}\)^2+\cdots\)\varepsilon_f.
\eea

By keeping up to the lowest order of
${k_f}/{m}$, the ratio of the binding
    energy to the Fermi kinetic energy is $\0{4}{15}$. Furthermore, one can obtain the
    the universal dimensionless coefficient $\xi$, i.e, the ratio of the energy density to that of a free Fermi
    gas
\bea \xi=\0EN/{\(\0EN\)_{free}}=\04{15}\times \053=\049. \eea
    This analytical result is in the range of the existed theoretical ones
    $\xi\sim 0.3-0.6$\refr{Heiselberg2000,physics/0303094,Ho2004,Astrakharchik2004,Astrakharchik2005,Schwenk2005,Carlson2005,Cohen2005,Dean2006,Chevy2006,Hu2006,Nishida2006,Bhattacharyya2006,Rupak2006,cond-mat/0404687,Bulgac2005}.
It is exactly consistent with that of the quantum
    Monte Carlo calculations\refr{physics/0303094} and in reasonable
    agreement with the updating experiments\refr{Bartenstein2004,Kinast2005,Partridge2006,Stewart2006}.
    We also note a similar $\xi _{D=\infty}=\049$ was obtained by
    Steele within the effective theory long ago, with $D$ being the space-time dimensions\refr{Steele}.

At unitarity, the energy gap should be also proportional to the
Fermi kinetic energy according to the dimensional analysis. It can
be \textit{empiristicly} derived from the total energy density
with the usual odd-even staggering in the thermodynamics
limit\refr{physics/0303094,Cohen2005} $\Delta =\0{5}{18}\varp_f$,
with $\varp_f$ being the Fermi kinetic energy. The critical
 temperature is $T_c\approx 0.157
    T_F$ approximated with the BCS universal relation $T_c={e^{\gamma}\Delta}/{\pi
    }$, where $\gamma$ is the
    Euler constant. The $T_c$ is in agreement
    with the updating theoretical\refr{Bulgac2005,Burovski2006}/experimental\refr{Kinast2005,Luo2007} results.
It should be pointed out that the differences for the energy gap
$\Delta$/critical BCS phase transition temperature $T_c$ can be as
large as several times in the literature. In the meantime, the
validity of the weak coupling BCS relation between $\Delta $ and
$T_c$ deserves to be further studied theoretically.

Let us further analyze the pressure by keeping up only to the
lowest order
    of $k_f$
\bea P=\01{30m}\(3\pi^2\)^{\023}n^{1+\023}=\alpha
    n^{1+\023}.
\eea The power index of $n$ is also consistent with the existed
    theoretical calculations\refr{Heiselberg2004}.
One can see that the strongly correlating effect as well as the
    fermion mass affects the coefficient $\alpha$ significantly.

The sound speed squared for the ideal Fermi gas is \bea
    v_{FG}^2=\01m\0{\pa P_{FG}}{\pa n}=\013v_f^2,
\eea
    with the Fermi velocity $v_f={k_f}/{m}$.
In the unitary
    limit,
    the pressure is $P=\016P_{FG}$,
    from which one can find the sound speed is reduced remarkably
\bea
    v=\sqrt{\016}v_{FG}=\0{\sqrt{2}}6v_f>0.
\eea It has been
    argued that the sound speed squared might be negative due to the
    theoretical spinodal instability in the unitary
    limit\refr{Heiselberg2000}.
Although the sound speed is significantly reduced
    due to the correlating effects,
    it is still a real number which indicates the system
    stability.

The \textit{interior} correlations have been refreshingly
incorporated
    in the thermodynamics as an \textit{external} source manner indirectly
    through a mirror Thomson background.
In other words, we have derived the effective interaction strength,
with which we have given the general but concise analytical
thermodynamic expressions at both finite temperature and density.
This analytic method can be easily extended to the near unitary
limit regime\refr{chen2006}.

In conclusion, there is a similarity between the strongly
interacting ultra-cold atoms physics in the scattering length
    limit $|a|\to \infty$ and the infinite
instantaneous Coulomb interaction in a compact confinement
environment due to the competition of long range forces with short
range ones. The classical Thomson Problem as a potential quantum
many-body arm redounds to addressing the universal thermodynamics,
with which the key coefficient $\xi
    =\0{4}{9}$ and the energy gap $\Delta =\0{5}{18}\varp_f$ are obtained  in the QED-like framework
    with HLS formalism.

{\bf Acknowledgments}: The author thanks Professor J. R. Li for
discussion. Supported in part by the Central China Normal
University scientific research fund and NSFC under grant No
10675052.

\end{document}